\documentclass[twocolumn,aps,amsmath,amssymb,floatfix]{revtex4}

\usepackage{bm}
\begin{document}
\title{Bulk and Boundary Critical Behavior at Lifshitz
  Points\footnote{Invited talk presented at the 22nd International
    Conference on Statistical Physics (STATPHYS 22) of the
    International Union of Pure and Applied Physics (IUPAP), 4--9 July
  2004, Bangalore, India, to be published in {\it PRAMANA---Journal of
  Physics}.}}
\author{H.~W. Diehl}

\affiliation{Fachbereich Physik, Universit{\"a}t Duisburg-Essen, Campus Essen, D-45117 Essen, Germany}

\begin{abstract}Lifshitz points are multicritical points at which a disordered
  phase, a homogeneous ordered phase, and a modulated ordered phase meet.
  Their bulk universality classes are described by natural
  generalizations of the standard $\phi^4$ model. Analyzing these
  models systematically via modern field-theoretic renormalization
  group methods has been a long-standing challenge ever since their
  introduction in the middle of the 1970s. We survey the recent
  progress made in this direction, discussing results obtained via
  dimensionality expansions, how they compare with Monte
  Carlo results, and open problems. These advances opened the way
  towards systematic studies of boundary critical behavior at
  $m$-axial Lifshitz points. The possible boundary critical behavior
  depends on whether the surface plane is perpendicular to one of the
  $m$ modulation axes or parallel to all of them. We show that the
  semi-infinite field theories representing the corresponding surface
  universality classes in these two cases of perpendicular and
  parallel surface orientation differ crucially in their Hamiltonian's
  boundary terms and the implied boundary conditions, and explain
  recent results along with our current understanding of this matter.
\end{abstract}
\keywords{Critical behavior, Lifshitz points, field theory, boundary
  critical behavior, anisotropic scale invariance}

\pacs{05.20.-y, 11.10.Kk, 64.60.Ak, 64.60.Fr, 05.70.Np}

\maketitle

\section{Introduction}

Lifshitz points (LP) are a particular kind of multicritical points at
which a disordered phase meets both a spatially homogeneous ordered
phase as well as a modulated ordered one
\cite{Hor80,Sel88,Sel92,Die02}.  They were introduced in 1975 by
Hornreich, Luban, and Shtrikman \cite{HLS75a}, though apparently
discovered independently by two other groups \cite{DC-F75KO76} (cf.\ 
Ref.~\cite[p.~59]{Sel92}). Their discovery triggered considerable
theoretical
\cite{HUB76,AD77,Mic77,RS77,GS78,HB78,HLS78,ML78,SG78,HLS79,HLSS79,Hor79,AM80,Sel7880,Tal80,YCS84,Oit85,MF91}
and experimental interest \cite{BSOC8081,MZBRW82}, which has continued
over the years, and after a phase of somewhat reduced intensity, has
regained a lot of momentum recently
\cite{BMLSM95,SMFA9900,BBO00,SBMRR00WEHFB00,VS92,LJS00,LS-G02}---in
particular, on the theory side
\cite{FM93,FH93,Bar95,NF95,DL96,KM97,NAF97,MC9899,ZSK00,DS00a,%
  Lei00,AL01,SD01,DS01a,DS02,DS03,Lei03,DSZ03}.

The physics of LP is interesting for a variety of reasons. Let me
mention a few.
\begin{list}{(\roman{enumi})}{\usecounter{enumi}}
\item A \emph{wealth of physically distinct systems} exist that are
  either known to have LP or for which LP have been discussed; this
  includes such diverse systems as magnets \cite{BSOC8081},
  ferroelectrics \cite{VS92}, polymer mixtures
  \cite{DL96,SMFA9900,MS04}, liquid crystals \cite{CL76}, systems
  undergoing structural phase transitions or domain wall instabilities
  \cite{LN79}, organic crystals \cite{AD77,HZW80}, and even
  superconductors \cite{Nog00}.
\item The physics of LP embodies \emph{many of the crucial concepts}
  of the modern theory of phase transitions and critical phenomena,
  yet has been explored \emph{to a much lesser degree} than critical
  behavior at conventional critical points. The best studied
  universality classes of bulk critical behavior are the ones for
  $d$-dimensional systems with short-range interactions and an
  $n$-component order parameter field $\bm{\phi}$, represented
  by the $O(n)$ $\phi^4$ models.  For them, very detailed---and in
  part impressively accurate---results have been worked out by means
  of sophisticated renormalization group (RG) approaches
  \cite{ZJ96,PV02}, series expansions \cite{BC96}, and computer
  simulations \cite{Lan99}; and many of these theoretical predictions
  have been checked by careful experiments.
  
  By contrast, the application of modern field-theoretic RG approaches
  to the study of critical behavior at LP is a fairly recent
  development \cite{MC9899,DS00a,SD01,DSZ03}. The two-loop RG analysis
  of critical behavior at $m$-axial LP in $d=4+\frac{m}{2}-\epsilon$
  dimensional systems Shpot and myself \cite{DS00a,SD01} managed to
  perform for general values $0\leq m\leq d$ has finally yielded the
  $\epsilon$~expansions of all critical exponents to second order. The
  estimates obtained by means of these series expansions for the
  values of the critical exponents for the scalar uniaxial case
  ${n=m=1}$ in ${d=3}$ dimensions agree quite well with up-to-date
  Monte Carlo results \cite{PH01}.  Unfortunately, we are aware of
  only a few high-temperature series estimates \cite{RS77,Oit85,MF91},
  none of which is very recent. On the experimental side, renewed
  activity is noticeable. Aside from the recent work on polymer
  mixtures \cite{BMLSM95,SMFA9900}, new experiments on magnetic
  systems have been reported \cite{BBO00,SBMRR00WEHFB00}. However, so
  far the latter have not produced results for the critical and
  crossover exponents of the $m=n=1$ LP point of significantly greater
  accuracy than achieved in previous studies \cite{BSOC8081,ZSK00}.
\item Compared with critical points, LP provide \emph{additional
    challenges}. Since they are multicritical points, a further
  thermodynamic variable besides temperature $T$ must be fine-tuned to
  reach them. Furthermore, precise experimental investigations of
  their critical behavior should include verification of the expected
  crossover scaling forms and expected to involve the choice of proper
  nonlinear scaling fields \cite{DSZ03,DGR03}.
  
  On the theoretical side, progress in analytical RG analyses has been
  hampered by the \emph{substantial technical difficulties} one
  encounters in computations of Feynman diagrams beyond one-loop
  order.  The origins of these problems are twofold: the anisotropic
  nature of scale invariance that holds at the LP, which implies that
  the free propagator does not reduce to a simple power at the LP but
  involves a scaling function; and the fact that this scaling function
  in position space turns out to have a rather complicated form in
  general \cite{DS00a}. The progress made recently \cite{DS00a,SD01}
  in handling such field theories could pave the way for systematic
  investigations of \emph{general aspects of anisotropic scale
    invariance} (ASI) in systems with short-range interactions. One
  important question that has been raised long ago \cite{Car85,Hen97}
  but not yet answered in a truly convincing fashion is the following.
  Scale invariance, in conjunction with translation and rotation
  invariance, and short range of interactions, is known to normally
  imply invariance under a larger symmetry group, namely under
  conformal transformations \cite{Pol70,Car87,Pol88}. Does ASI
  likewise entail the invariance under additional nontrivial
  continuous transformations? Henkel has played with this idea for
  years \cite{Hen97,Hen9902}; making concrete propositions for
  transformations under which two-point correlation functions should
  be invariant, he has come up with definite predictions for the form
  of the associated scaling functions, which appear to be consistent
  with Monte Carlo results \cite{PH01} for the three-dimensional ANNNI
  model, yet remain to be carefully checked by analytical calculations
  \cite{aging,CG02b}. The field theories representing the universality
  classes of critical behavior at $m$-axial LP are particularly well
  suited for such scrutiny, not least because the parameter $m$ can be
  varied.
\item Since LP involve both modulated ordered phases as well as ASI,
  rich and interesting \emph{boundary critical phenomena}
  \cite{Bin83,Die86a,Ple04} may be expected to occur near them. The
  systematic investigation of such phenomena, in particular, via
  field-theoretic RG tools, is still in its infancy
  \cite{DGR03,Gum86,BF99,FKB00,Ple02,DR04}.
\end{list}

In this contribution, I will briefly survey the progress made recently
in the application of field-theoretic RG methods to bulk and boundary
critical phenomena at LP, compare its results with those from other
sources such as Monte Carlo simulations, highlight some of the central
issues and difficulties, and indicate directions for further research.
We begin in the next section by specifying the models, then deal with
their bulk critical behavior, before we turn in Sec.~\ref{sec:bcb} to
the issue of boundary critical behavior.

\section{Continuum models and bulk critical behavior}
\label{sec:cbm}

\subsection{Continuum models}
\label{sec:cm}

Having in mind systems whose microscopic interactions are either short
ranged or of a long-range kind that is irrelevant in the RG sense, we
consider continuum models with a Hamiltonian of the form
\begin{equation}
  \label{eq:Hamfb}
    {\mathcal{H}}={\int_{\mathfrak{V}}}{\mathcal L}_{\rm b}(\bm{x})\,
    {\rm d} V+ {\int_{\mathfrak{B}}}{\mathcal L}_1(\bm{x})
    \,{{\rm d}}A\;, 
\end{equation}
where ${\mathcal L}_{\rm b}(\bm{x})$ and ${\mathcal
  L}_1(\bm{x})$ are functions of the $n$-component order
parameter $\bm{\phi}(\bm{x})=(\phi_a(\bm{x}))$
and its derivatives with respect to the coordinates
$(x_\alpha,x_\beta)\equiv\bm{x}$. We index the first $m$
Cartesian coordinates by $\alpha$; they refer to the $m$-dimensional
subspace to which the modulation of order is confined.  The remaining
$\bar{m}\equiv d-m$ ones are labeled by $\beta$. When we deal with
boundary critical behavior, the volume and surface integrals
$\int_{\mathfrak{V}}{\rm d} V$ and $\int_{\mathfrak{B}}{\rm d} A$
extend over the half-space ${\mathbb
  R}^d_+=\{\bm{x}=(\bm{r},z)|\bm{r}\in{\mathbb
  R}^{d-1},{0\leq z<\infty}\}$ and the $z=0$ hyperplane
$\mathfrak{B}$, respectively.  To investigate bulk critical behavior,
we may as well take ${\mathfrak{V}}={\mathbb R}^d$ and forget about
the boundary piece in Eq.~(\ref{eq:Hamfb}), choosing appropriate
(periodic) boundary conditions.
\begin{widetext}
Unless stated otherwise, the bulk density is
\begin{eqnarray}
  \label{eq:Lb}
 {\mathcal L}_{\rm b}(\bm{x})&=&\frac{\mathring{\sigma}}{2}\,
  \bigg(\sum_{\alpha=1}^m\partial_\alpha^2\bm{\phi} \bigg)^2
 +\frac{1}{2}\,
\sum_{\beta=m+1}^d{(\partial_\beta\bm{\phi})}^2
+\frac{\mathring{\rho}}{2}\,
\sum_{\alpha=1}^m{(\partial_\alpha\bm{\phi})}^2
 +\frac{\mathring{\tau}}{2}\,
\bm{\phi}^2+\frac{\mathring{u}}{4!}\,|\bm{\phi} |^4
\end{eqnarray}
in the sequel. Here $\partial_\alpha\equiv\partial/\partial x_\alpha$
and $\partial_\beta\equiv \partial/\partial x_\beta$, and
$\mathring{\sigma}>0$ as well as $\mathring{u}>0$ is assumed. For the
time being we focus on \emph{bulk} critical behavior. Let us therefore
postpone the choice of the boundary density ${\mathcal L}_1$ to
Sec.~\ref{sec:bcb} Our selection (\ref{eq:Lb}) of ${\mathcal L}_{\rm
  b}$ reflects two tacitly assumed properties: $O(n)$ invariance, and
isotropy in the $m$-dimensional $\alpha$-subspace of coordinates. An
investigation of the effects of spin anisotropies breaking the $O(n)$
invariance of ${\mathcal L}_{\rm b}$ may be found in
Ref.~\cite{Hor79}; they will not be considered here. However, the role
of ``space anisotropies'' reducing the Euclidean invariance in the
$\alpha$-subspace \cite{DSZ03} will be briefly discussed at the end of
this section.
\end{widetext}

From Eq.~(\ref{eq:Lb}) it is easy to understand how a LP can occur.
The interactions constants $\mathring{\sigma},\ldots,\mathring{u}$ all
depend on $T$ and a second thermodynamic variable, a non-ordering
field $g$ \cite{Die02} such as pressure (charge-transfer salts
\cite{AD77,HZW80}), a ratio of next-nearest neighbor (nnn)
antiferromagnetic and nearest-neighbor (nn) ferromagnetic interactions
along an axis (ANNNI model \cite{Sel88}), or a magnetic field
component in the subspace orthogonal to the order parameter (the
orthorhombic magnetic crystal MnP \cite{BSOC8081,YCS84,BBO00,ZSK00}).
Assuming that the coefficient of the
$(\partial_\beta\bm{\phi})^2$ term does not change sign, we
have absorbed it in the amplitude of $\bm{\phi}$ so that it
becomes $1/2$. Landau theory gives a disordered phase for
$\mathring{\tau}>0$ provided $\mathring{\rho}>0$, separated from a
homogeneous ordered one by the critical line
$\mathring{\tau}_c(\mathring{\rho}\geq 0)=0$. For negative
$\mathring{\rho}$, a continuous transition from the disordered to a
modulated ordered phase occurs across the so-called ``helicoidal
section'' $\tau=\mathring{\tau}_c(\mathring{\rho}<0)$ of the critical
line, which joins the ``ferromagnetic section'' at the LP
${\mathring{\tau}=\mathring{\rho}=0}$ (see, e.g.\ Fig.~1 of
Ref.~\cite{Die02}). The other phase boundary emerging from the LP
separates the homogeneous ordered from the modulated ordered phase.
The transitions across it can be of first or second order; for cases
with a scalar order parameter they are generically discontinuous,
whereas for specific models with a vector order parameter they turn
out to be continuous \cite{CL95}.

\subsection{Critical exponents, anisotropic scale invariance}
\label{sec:critexp}

In Landau theory, the helicoidal section $\mathring{\tau}_{\rm
  hc}\equiv \mathring{\tau}_c(\mathring{\rho}<0)$ varies as
$\mathring{\tau}_{\rm hc}\sim \mathring{\rho}^2$ near the {LP}. Beyond
Landau theory, the LP and the phase boundaries---supposing they still
exist---get shifted as a result of fluctuations, and the helicoidal
section of the critical line is expected to behave near the LP as
\begin{equation}
  \label{eq:codef}
  \mathring{\tau}_{\rm hc}\left(\delta\mathring{\rho}\right)
  -\mathring{\tau}_{\rm LP} \equiv \delta\mathring{\tau}_{\rm hc}
  \sim\left|\delta\mathring{\rho}\right|^{1/\varphi}\sim|\delta
  g|^{1/\varphi}\;.
\end{equation}
Here $\delta\mathring{\rho}$ and $\delta g$ denote deviations of
$\mathring{\rho}$ and $g$ from their values $\mathring{\rho}_{\rm LP}$
and $g_{\rm LP}$ at the {LP}. We have introduced the crossover
exponent $\varphi$, whose mean-field value is $\varphi_{\rm MF}=1/2$,
and utilized the fact that $\delta\mathring{\rho}\sim \delta g$ near
the {LP}.

In the modulated ordered phase, the order is modulated with a wave
vector $\bm{q}_{\rm mod}(T,g)$ depending on $T$ and $g$. Since
homogeneous order corresponds to $\bm{q}_{\rm mod}=0$,
$\bm{q}_{\rm mod}$ must also vanish at the {LP}. Its limiting
behavior as the LP is approached along the critical line's helicoidal
section $T=T_{\rm hc}(g)$ is governed by the wave-vector exponent
$\beta_q$, defined via
\begin{equation}
  \label{eq:wvexp}
  {q}_{\rm mod}(T_{\rm hc},g)\sim \left| \delta\mathring{\rho}\right|^{\beta_q}
  \sim |\delta g|^{\beta_q}\;.
\end{equation}

Other important critical exponents characterize the scale invariance
at the {LP}. Let us set
${\mathring{\tau}=\mathring{\rho}=\mathring{u}=0}$ in
Eq.~(\ref{eq:Lb}) and transform to momentum ($\bm{q}$) space
to obtain the two-point bulk vertex function $ \Gamma_{\rm
  b}^{(2)}(\bm{q})$ in the Ornstein-Zernicke approximation.
Its $\bm{q}$-dependence reads $\mathring{\sigma}\,(q_\alpha
q_\alpha)^2 +q_\beta q_\beta$, where repeated indices $\alpha$ and
$\beta$ are to be summed over $1,\ldots,m$ and $m+1,\ldots,d$,
respectively. Beyond this classical approximation one anticipates
again nontrivial power laws. Hence one introduces analogs of the usual
correlation exponent $\eta$ by
\begin{equation}
  \label{eq:etadefs}
  \Gamma_{\rm b}^{(2)}(\bm{q})\mathop{\sim}_{\bm{q}\to \bm{0}}
\left\{  
\begin{array}[c]{ll}
   q^{2-\eta_{L2}}& \text{ for } q_\alpha =0,\\[\smallskipamount]
   q^{4-\eta_{L4}}& \text{ for } q_\beta =0.
  \end{array}\right.
\end{equation}
These relations mean that $q_\alpha$ scales as $q_\alpha \sim
(q_\beta)^\theta$, with the ``anisotropy exponent''
\begin{equation}
  \label{eq:thetaeta}
  \theta =(2-\eta_{L2})/(4-\eta_{L4})\;;
\end{equation}
in Landau theory it takes the value $\theta_{\rm MF}=1/2$. Likewise
$x_\alpha\sim (x_\beta)^\theta$.

To formulate ASI in position space, let us consider a perturbation
$g_{\mathcal O}\int_{\mathfrak{V}}{\mathcal O}(\bm{x})\,{\rm
  d} V$ of the fixed-point Hamiltonian associated with the LP, where
${\mathcal O}(\bm{x})$ is a scaling operator with scaling
dimension $\Delta[{\mathcal O}]$. Let $y_{\mathcal O}$ be the RG
eigenexponent of the associated scaling field $g_{\mathcal O}$, so
that $g_{\mathcal O}\to \bar{g}_{\mathcal O}(\ell)=\ell^{-y_{\mathcal
    O}}\,g_{\mathcal O}$ under scale transformations $x_\beta\to
x_\beta \,\ell$. Since the scaling dimension $\Delta[{\mathcal O}]$
and the eigenexponent $y_{\mathcal O}$ must add up to minus the scaling
dimension of the volume $V=\int_{\mathfrak{V}}{\rm d} V$, which is
$\bar{m}+m\,\theta$, we have
\begin{equation}
  \label{eq:yDel}
  y_{\mathcal O}=\bar{m}+m\,\theta-\Delta[{\mathcal O}]\;.
\end{equation}
The operators ${\mathcal O}(\bm{x})$ satisfy
\begin{equation}
  \label{eq:ASIO}
  {\mathcal O}(\ell^\theta x_\alpha,\ell\,
  x_\beta)=\ell^{-\Delta[{\mathcal O}]}\,{\mathcal O}(x_\alpha,x_\beta)
\end{equation}
(ASI) in the long-scale limit $\ell\to 0$. 

\subsection{Field theory and $\epsilon$ expansion}
\label{sec:epsexp}

For a conventional critical point it is known that below the upper
critical dimension $d^*=4$, where hyperscaling is valid, two
independent critical exponents exist in terms of which all critical
indices characterizing the leading infrared singularities can be
expressed. They derive from the scaling dimensions of $\bm{\phi}$ and
the energy density $\phi^2$, or equivalently, the RG eigenexponents $y_h$
and $y_\tau$. Furthermore, there are just two metric factors, one
associated with each of the corresponding scaling fields $h$ and $\tau$
(``two-scale factor universality'' \cite{SFW72TF75}). In the case of
$m$-axial LP, the upper critical dimension is \cite{HLS75a}
\begin{equation}
  \label{eq:UCD}
  d^*(m)=4+\case{m}{2}\;.
\end{equation}
The easiest way to see this is to determine the dimension $d=d^*(m)$
below which the Gaussian scaling dimension of $\mathring{u}$ becomes
positive; the Ginzburg criterion yields the same result.

In view of the different scaling of $x_\alpha$ and $x_\beta$, and the
need to fine-tune an additional variable---$\mathring{\rho}$ or
$g$---, it is natural to expect that four critical exponents will be
required to express the bulk critical exponents of the LP for
$d<d^*(m)$. Of course, some of these might turn out to be trivial,
taking on values independent of $d$ and $m$. For example, one might
anticipate the anisotropy exponent $\theta$ to retain its mean-field
value for $d<d^*(m)$.  However, the $\epsilon$-expansion results of
Shpot and myself \cite{DS00a,SD01} have revealed that nontrivial
$m$-dependent contributions to $\theta$ appear at order $\epsilon^2$.
The bulk operators ${\mathcal O}(\bm{x})$ from whose scaling
dimensions the four independent bulk exponents derive are given in
Table~\ref{Tab:bsop}, along with the associated scaling fields and
their RG eigenexponents.  Each of these four scaling fields involves a
nonuniversal metric factor. Hence a four-scale-factor universality
applies.

\begin{table*}[t]
\caption{Bulk scaling operators ${\mathcal O}(\bm{x})$, associated
  scaling dimensions $\Delta[{\mathcal O}]$, bulk scaling fields
  $g_{\mathcal O}$, and their RG
  eigenexponents $y_{\mathcal O}$, giving the four independent bulk
  critical exponents of the {LP}. \label{Tab:bsop}}
\begin{ruledtabular}
\begin{tabular}{cccc}
 ${\mathcal O}(\bm{x})$&$\Delta[{\mathcal O}]$&$g_{\mathcal O}$ &$y_{\mathcal
   O}$\\ \hline
$\bm{\phi}$&$(\bar{m}+m \theta-2+\eta_{L2})/2$&$\bm{h}$&$(\bar{m}+m\theta+2-\eta_{L2})/2$\\ 
$(\partial_\alpha\partial_\alpha\bm{\phi})^2$&$ \bar{m}+m\,\theta-4\theta+2 $&$\sigma$&$4\theta-2$\\
$\phi^2$&$\bar{m}+m\theta-1/{\nu_{L2}}=(1-\alpha_L)/{\nu_{L2}}$&$\tau$&${1}/{\nu_{L2}}$\\
$(\partial_\alpha \bm{\phi})\partial_\alpha \bm{\phi}$&$\bar{m}+m\theta-\varphi /{\nu_{L2}}$&$\rho$&$\varphi/ \nu_{L2}$\\
\end{tabular}
\end{ruledtabular}
\end{table*}

Given a line of upper critical dimensions $d^*(m)$, one should be able
to expand about any point on it. Although this goal was identified at
a very early stage \cite{HLS75a}, its implementation turned out to be
very demanding and took a long time. In Refs.~\cite{DS00a,SD01} a
two-loop RG analysis was performed in $d^*(m)-\epsilon$ dimensions for
general values of $m$. This gave the $\epsilon$~expansions of the four
independent bulk critical exponents $\eta_{L2}$, $\theta$, $\nu_{L2}$,
and $\varphi$, as well as the correction-to-scaling exponent
$\omega_u$, to $O(\epsilon^2)$.

\begin{widetext}
Technically, a massless minimal-subtraction renormalization scheme was
employed. To define the ultraviolet (uv) finite renormalized theory,
the reparametrizations
\begin{eqnarray}
  \label{eq:bulkrep}
  \bm{\phi}&=&Z_\phi^{1/2}\,\bm{\phi}_{\text{ren}}\;,\qquad
  \mathring{\sigma}=Z_\sigma\,\sigma\;,\qquad
\mathring{u}\,{\mathring{\sigma}}^{-m/4}\,F_{m,\epsilon}=
   \mu^\epsilon\,Z_u\,u\;,
\nonumber\\
\mathring{\tau}-\mathring{\tau}_{\text{LP}}&=&
\mu^2\,Z_\tau\,{\big[\tau+A_\tau\,\rho^2\big]}\;,\qquad 
\left(\mathring{\rho}-\mathring{\rho}_{\text{LP}}\right)\,
{\mathring{\sigma}}^{-1/2}=\mu\,Z_\rho\,\rho\;.
\end{eqnarray}
were made, where $\mu$ is a momentum scale, while $F_{m,\epsilon}$ is
a convenient (uv finite) normalization factor whose precise choice
need not worry us here. All renormalization factors $Z_\phi$,
$Z_\sigma$, $Z_\tau$, $Z_\rho$, and $Z_u$ were computed to $O(u^2)$.
From the result for $Z_u$ the RG beta function $\beta_u(u,\epsilon)$
follows to order $u^3$; the other $Z$-factors yield RG functions whose
values at the nontrivial root $u^*(m,\epsilon)$ of $\beta_u$ give the
critical exponents.  The main consequence of the renormalization
function $A_\tau$ is that the scaling field with the RG eigenexponent
$1/\nu_{L2}$ becomes a linear combination of $\tau$ and $\rho^2$
\cite{DSZ03,DGR03}.

What makes calculations beyond one-loop order complicated is that the
scaling function $\Phi_{m,d}(\upsilon)$ of the free bulk propagator at
the LP,
\begin{eqnarray}
  \label{eq:Gb}
  G_{\rm b}(|\bm{x}|)&=&\int \frac{{\rm d}^dq}{(2\pi)^{d/2}}\,
  \frac{\exp(i\bm{q}\cdot\bm{x})}{q_\beta
    q_\beta+\mathring{\sigma}(q_\alpha q_\alpha)^2} 
\nonumber\\ &=& 
\mathring{\sigma}^{-m/4}\,(x_\beta x_\beta)^{-1+\epsilon/2}\,
   \Phi_{m,d}(\upsilon)\;,
\quad \upsilon\equiv{\big(\mathring{\sigma}x_\beta x_\beta\big)}^{-1/4}
\sqrt{x_\alpha x_\alpha} \;,
\end{eqnarray}
is a difference of generalized hypergeometric functions. While these
increase in general exponentially as ${\upsilon\to\infty}$, their
difference has an asymptotic expansion in inverse powers of $\upsilon$
that does not terminate except for special choices of $(m,d)$, such as
$(2,5)$ and $(6,7)$, where it reduces to elementary functions
\cite{DS00a,SD01}.  Therefore, the two-loop series coefficients of the
renormalization factors could not be computed analytically for general
$m$. However, they---as well as the implied $\epsilon$-expansion
coefficients of the critical exponents---could be written in terms of
four single integrals $j_\phi(m)$, $j_\sigma(m)$, $j_\rho(m)$, and
$j_u(m)$ of the form $\int_0^\infty{\rm d} \upsilon\,f(\upsilon;m)$,
where $f(\upsilon;m)$ involves $\Phi_{m,d^*(m)}(\upsilon)$, analogous
(related) scaling functions, and powers of $\upsilon$ \cite{SD01}. For
$m=0,\,2,\,6,\,8$, these integrals could be computed analytically; for
other values of $m$ they had to be determined by numerical means.
\end{widetext}

The resulting $\epsilon$~expansions of the critical exponents
$\lambda=\nu_{L2}$, \ldots, $\varphi $ and the correction-to-scaling
exponent $\omega_u$ take the form
\begin{equation}
  \label{eq:lamexp}
  \lambda(n,m,d)=\lambda_{\rm MF} +\lambda_1(n)\,\epsilon +\lambda_2(n,m)\,\epsilon^2 +O(\epsilon^3) \;.
\end{equation}
Note that $\lambda_1(n)$ is \emph{independent of} $m$, so that the
$m$-dependence starts at order $\epsilon^2$. This means that the
coefficients $\lambda_1(n)$ coincide with their $m=0$ counterparts for
the standard $\phi^4$ model for all exponents that remain meaningful
when $m=0$.  (Recall that exponents such as $\eta_{L4}$, $\varphi$,
and $\theta$ are not needed in the isotropic case $m=0$.)

The result (\ref{eq:lamexp}) allows several interesting checks. First,
if we substitute the analytically known ${m=0}$ values of the
integrals $j_\iota(m)$ into it, choosing $\lambda=\eta_{L2}$,
$\nu_{L2}$, and $\omega_u$, then the familiar expansions to
$O(\epsilon^2)$ of the standard exponents $\eta$, $\nu$, and
$\omega_u$ of the $\phi^4$ model are recovered. A second check
concerns the special cases ${m=2}$ and ${m=6}$. Owing to enormous
simplifications, the two-loop RG analysis can be performed fully
analytically. The results one obtains in this fashion are fully
consistent with what one gets upon insertion of the analytically known
values of $j_\iota(2)$ and $j_\iota(6)$ into the two-loop expressions
for general $m$. Third, considering the case of the isotropic LP
\cite{HLS75a,DS02}, one can set ${d=m=8-\epsilon_8}$ in
Eq.~(\ref{eq:lamexp}) and expand to second order in
$\epsilon_8=2\epsilon$. The limiting values $j_\iota(8-)$ are again
known analytically \cite{SD01,DS02}. Considering exponents that remain
meaningful in the isotropic case $m=d$, such as $\eta_{L4}$ or
$\nu_{L4}=\theta\,\nu_{L2}$, we can derive their expansions in
$\epsilon_8$ to $O({\epsilon_8}^2)$ from Eq.~(\ref{eq:lamexp}). The
results agree with those obtained via a direct analysis of the
isotropic model with $m=d$ in $8-\epsilon_8$ dimensions
\cite{HLS75a,DS02}.

A cautionary remark is appropriate here. As a candidate for an
experimental system with an isotropic Lifshitz point, ternary mixtures
of A and B homopolymers and AB diblock copolymers have been studied
both experimentally \cite{SMFA9900} and theoretically \cite{MS04}. In
their case, modulated order occurs in the lamellar phase. While
self-consistent field theory predicts the transition from the
disordered to the lamellar phase to be continuous \cite{MS04},
theoretical arguments in favor of a first-order transition have been
presented \cite{Bra75FH87}. This would mean that there is actually no
isotropic {LP}. According to some experiments (see the discussion in
Sec.~7 of Ref.~\cite{MS04}), the Lifshitz point found in mean-field
(MF) theories gets apparently destroyed. Unfortunately, recent Monte
Carlo simulations \cite{MS04} were not able to decide whether the
transition between the disordered and lamellar phases is of first
order or continuous. However, they yielded modifications of the MF
phase diagram similar to those seen in experiments---in particular, no
{LP}. If fluctuations indeed preclude the appearance of an isotropic
LP, then the $\epsilon$~expansions of the critical exponents for that
case are mainly of academic interest. Nevertheless, their consistency
with the results for general $m$ is very gratifying from a
mathematical point of view.

The series-expansion results for general $m$ can be, and were, used
in particular to obtain approximate values for the critical exponents
of the uniaxial LP with ${n=1}$ at ${d=3}$ \cite{SD01}. Both
experiments on MnP \cite{BSOC8081} well as Monte Carlo calculations
for the ANNNI \cite{Sel7880,PH01} model provide clear evidence for the
existence of such a {LP}. Recent field-theoretic estimates are
$\nu_{L2}\simeq 0.75$, $\beta_L\equiv \nu_{L2}\,\Delta[\phi]\simeq
0.22$, $\theta=\nu_{L4}/\nu_{L2}\simeq 0.47$, $\varphi\simeq 0.68$,
$\alpha_L\simeq 0.16$, and $\gamma_L\simeq 1.4$ \cite{SD01}. The
agreement with current Monte Carlo results, which gave
$\alpha_L=0.18\pm 0.03$, $\beta_L=0.235\pm 0.005$, and
$\gamma_L=1.36\pm 0.03$, is fairly good. For more detailed comparisons
covering also other cases, experimental work, and further theoretical
estimates the reader is referred to Refs.~\cite{SD01,PH01,Ple04}.

\begin{widetext}
\subsection{Space anisotropies}
\label{sec:spais}

A natural generalization of the ANNNI model is the biaxial nnn Ising
(BNNNI) model, which has competing nn and nnn interactions along two
cubic axes rather than along a single one. In $d$ dimensions, even
$m$-axial variants of the latter, ``mNNNI models'' with $m\leq d$, can
be considered. The continuum models onto which they map upon coarse
graining generically have fourth-order derivative terms \emph{breaking
  isotropy} in the $\alpha$-subspace. Their symmetry may be cubic
or---if we consider similar microscopic systems involving other
crystal lattices---even weaker. Hence, whenever $m>1$, the bulk
density (\ref{eq:Lb}) should be supplemented by anisotropic
contributions of the form
\begin{equation}
  \label{eq:spaniso}
  {\mathcal L}_{\rm b}^{(w)}=\frac{\mathring{\sigma}}{2}\,\mathring{w}_i
  \,{T^{(i)}_{\alpha_1\alpha_2\alpha_3\alpha_4}}
  (\partial_{\alpha_1}\partial_{\alpha_2}\bm{\phi})
  \,\partial_{\alpha_3}\partial_{\alpha_4}\bm{\phi}=
  \frac{\mathring{\sigma}}{2}\mathring{w}\sum_{\alpha=1}^m
  (\partial_\alpha^2\bm{\phi})^2+\ldots,
\end{equation}
where all tensors $T^{(i)}_{\alpha_1\alpha_2\alpha_3\alpha_4}$
permitted by symmetry must be included. The $\mathring{w}_i$ are
dimensionless interaction constants. For cubic symmetry, only the
first term on the far right remains.
\end{widetext}

The effects of such space anisotropies were investigated in
Ref.~\cite{DSZ03}. A new renormalization factor $Z_{w_i}$ is required
for each independent anisotropy, and both these as well as the
previously introduced renormalization functions
[Eq.~(\ref{eq:bulkrep})] now depend on $u$ and the renormalized
anisotropies $w_i$. Specifically, the crossover exponent $\varphi_2$
associated with the cubic anisotropy $\mathring{w}$ was computed to
$O(\epsilon^2)$. For $m=2$ and $m=6$, the $O(\epsilon^2)$ coefficient
could be determined analytically, for other values of $m$ expressed in
terms of another (numerically computable) single integral. It turned
out to be small, but positive.  For example, for $m=2$ and $n=1$ its
evaluation at $\epsilon=2$ yielded the $d=3$ estimate $\varphi_2=
1/81\simeq 0.012$.  Thus the isotropic fixed point is $w_i=0$
unstable, at least for small $\epsilon$. Whenever such anisotropy is
present, the previously found universality classes should not apply.
Unfortunately, no new stable fixed point could be found. A detailed
clarification of the behavior for $w_i\neq 0$ remains a challenge. It
would be interesting to investigate the role of such anisotropies in
Monte Carlo simulations of suitably designed three-dimensional models
(e.g., the BNNNI model), albeit deviations from the $w_i=0$
universality classes may be difficult to measure because of the
smallness of $\varphi_2$.

\section{Boundary critical behavior at LP}
\label{sec:bcb}

The study of boundary critical behavior at LP started with Gumbs' work
based on Landau theory \cite{Gum86}, in which $z$ was taken to be an
$\alpha$-direction. Later considerably more detailed MF analyses
\cite{BF99,FKB00} and Monte Carlo calculations \cite{Ple02} of
semi-infinite ANNNI models with perpendicular ($z= \alpha$-direction)
and parallel ($z=\beta$-direction) surface orientations were
performed. So far, detailed field-theoretic RG studies were made only
for the case of parallel surface orientation \cite{DGR03,DR04}.

Let me emphasize that the two primary types of surface orientations
($\|$ or $\perp$) correspond to \emph{substantially distinct} cases.
This can be seen from the following observations: First, $z$ scales
differently, namely, as $\ell^{-1}$ and $\ell^{-\theta}$,
respectively. This has an immediate consequence. Consider a
perturbation $g_{{\mathcal O^{\mathfrak{B}}}}
{\int_{\mathfrak{B}}}{\mathcal O}^{\mathfrak B}(\bm{r})
\,{{\rm d}}A$, where ${\mathcal O}^{\mathfrak B}(\bm{r})$ is a
boundary operator with scaling dimension $\Delta[{\mathcal
  O}^{\mathfrak B}]$ and hence has the ASI property (\ref{eq:ASIO}).
The analogs of Eq.~(\ref{eq:yDel}) for the RG eigenexponent
$y_{{\mathcal O}^{\mathfrak B}}$ of $g_{{\mathcal O}^{\mathfrak B}}$
differ depending on the surface orientation:
\begin{eqnarray}
  \label{eq:gOBscl}
 y_{{\mathcal O}^{\mathfrak B}}&=&\bar{m}+m\,\theta
 -\Delta_{\|,\perp}[z]-\Delta[{\mathcal O}^{\mathfrak B}]\;, \nonumber\\[\medskipamount]
 \Delta_\|[z]&=&1\,,\quad\Delta_\perp=\theta\,.
\end{eqnarray}
Second, owing to the different engineering dimensions $[z]=\mu^{-1}$
and $[z]=\mathring{\sigma}^{1/4}\mu^{-1/2}$, power counting
considerations to estimate the relevance or irrelevance of
contributions to the surface density ${\mathcal L}_1$ differ. Third,
since Ginzburg-Landau theory yields differential equations for the
order parameter of second ($\|$) or fourth ($\perp$) order in
$\partial_z$, either a single or else two boundary conditions are
needed at ${z=0}$ and ${z=\infty}$.

To bring the problem into focus, let me recall that in the ${m=0}$
case of the standard semi-infinite $\phi^4$ model it is sufficient to
choose ${{\mathcal L}_1=\case12 \mathring{c}\,\phi^2}$, unless terms
breaking the $O(n)$ symmetry are permitted \cite{Die86a} (which will
be avoided here). On the basis of power counting alone, one might
think that the symmetry-allowed monomial
$\bm{\phi}\partial_n\bm{\phi}$ (where $\partial_n$
means derivative along the inner normal), should be included as well.
But this is \emph{redundant} because of the boundary condition
$\partial_n\bm{\phi} = \mathring{c}\,\bm{\phi}$, which
as usual follows from the boundary part of the classical equation
$\delta{\mathcal H}=0$ and holds beyond Landau theory inside of
averages.

The surface enhancement variable $\mathring{c}$ determines the type of
surface transition that occurs at bulk criticality: Depending on
whether its deviation
${\delta\mathring{c}}={\mathring{c}-\mathring{c}_{\rm sp}}$ from a
special value $\mathring{c}_{\rm sp}$ satisfies
${\delta\mathring{c}>0}$, ${\delta\mathring{c}=0}$ or
${\delta\mathring{c}<0}$ an ordinary, special or extraordinary
transition occurs \cite{Bin83,Die86a}, provided the dimension of the
surface, ${d-1}$, exceeds the value below which a long-range ordered
surface phase in the presence of a disordered bulk is not possible
(i.e., if $d>2$ and $d>3$ in the Ising and $n>1$ cases, respectively).
 
What modification occur in the $m>0$ LP case? They are easy to
understand if the surface orientation is parallel: An additional
derivative term must be included in ${\mathcal L}_1$, which thus
becomes \cite{DGR03}
\begin{equation}
  \label{eq:L1}
  {\mathcal L}^{\|}_1(\bm{x})=\frac{\mathring{c}}{2}\,
  \bm{\phi}^2 + 
\frac{\mathring{\lambda}}{2}\sum_{\alpha=1}^m
{(\partial_\alpha\bm{\phi})}^2\;.
\end{equation}
Since $[\mathring{c}] =[\mathring{\sigma}^{1/2}\partial_\alpha^2] =
\mu$, the variable $\mathring{\lambda}\sigma^{-1/2}$ is dimensionless.
The implied boundary condition reads
${(\partial_n-\mathring{\lambda}\partial_\alpha\partial_\alpha)}
\bm{\phi} =\mathring{c}\,\bm{\phi}$; it can be
employed to conclude that contributions to ${\mathcal L}_1^\|$ of the
form $\bm{\phi}\partial_n\bm{\phi}$ and
$(\partial_\alpha\bm{\phi})\partial_n\partial_\alpha\bm{\phi}$
are redundant. By contrast, the inclusion of the term
$\propto\mathring{\lambda}$ is necessary: Not only is it required to
absorb uv singularities of the theory, but it would be generated under
the RG if originally absent.  This can be seen as follows: In order to
renormalize the model defined by Eqs.~(\ref{eq:Hamfb}), (\ref{eq:Lb}),
and (\ref{eq:L1}), we must complement the reparametrizations
(\ref{eq:bulkrep}) by
\begin{eqnarray}
  \label{eq:surfrep}
  \bm{\phi}^{\mathfrak{B}}&=&(Z_\phi
  Z_1)^{1/2}\,\bm{\phi}^{\mathfrak{B}}_{\rm ren}\;,\nonumber\\
\mathring{\lambda}\,\mathring{\sigma}^{-1/2}&=&\lambda+
P_\lambda(u,\lambda,\epsilon)\;,\nonumber\\
 \mathring{c}-\mathring{c}_{\rm sp}&=&
  \mu\,Z_c{\big[c+A_c(u,\lambda,\epsilon)\,\rho\big]}\;.
\end{eqnarray}
Here the surface renormalization factors $Z_1$ and $Z_c$ depend on $u$
and $\lambda$, just as $P_\lambda$ and $A_c$. At $O(u^2)$, $P_\lambda$
does not vanish for ${\lambda=0}$, so a nonzero $\mathring{\lambda}$
gets indeed generated. Furthermore, there are no RG fixed points at
${\lambda=0}$ on the hyperplane $u=u^*$ (see Fig.~2 of
Ref.~\cite{DGR03}). The fixed points associated with the ordinary,
special, and extraordinary transitions turn out to be located at a
nontrivial $\lambda$-value $\lambda_+^*=\lambda_0(m)+O(\epsilon)$ and
$c={c^*_{\rm ord}\equiv\infty}$, $c^*_{\rm sp}\equiv 0$, and $c^*_{\rm
  ex}\equiv -\infty$, respectively.

Before continuing our account of the available results for this
parallel case, let us briefly discuss how to choose ${\mathcal L}_1$
when the surface orientation is perpendicular. Clearly, the two
monomials included in Eq.~(\ref{eq:L1}) should be expected here as
well, although different couplings ought to be associated with
$(\partial_\alpha\bm{\phi})^2$ for $\alpha=1$ ($z$-direction)
and $\alpha\geq 2$. As long as terms breaking the $O(n)$ symmetry can
be ruled out, the choice
\begin{equation}
  \label{eq:L1perp}
  {\mathcal L}^\perp_1=\frac{\mathring{c}_\perp}{2}\,\phi^2
  +\frac{\mathring{\lambda}_\|}{2}\sum_{\alpha=2}^{m}(
  \partial_\alpha\bm{\phi})^2  
  +\mathring{b}\,\bm{\phi}\partial_n\bm{\phi}
  +\frac{\mathring{\lambda}_\perp}{2}\,(\partial_n\bm{\phi})^2 
\end{equation}
should be sufficient. From the vanishing of the contributions
$\int_{\mathfrak B}\ldots\delta\partial_n\bm{\phi}$ and
$\int_{\mathfrak B}\ldots\delta\bm{\phi}$ to $\delta{\mathcal H}$
two boundary conditions on ${\mathfrak B}$ are found, namely
\begin{eqnarray}
  \label{eq:bcperp}
  {\left[\mathring{\sigma}\partial_n^3+(\mathring{b}-
  \mathring{\rho})\partial_n+\mathring{c}_\perp
  -\mathring{\lambda}_\|
  \sum_{\alpha=2}^m\partial_\alpha^2\right]}\bm{\phi} &=&0\;,
\nonumber\\[\medskipamount]
\left[-\mathring{\sigma}\,\partial_n^2+\mathring{\lambda}_\perp
  +\mathring{b}\right]\bm{\phi}&=&0\;.
\end{eqnarray}
They tell us that the monomials
$\bm{\phi}\partial_n^2\bm{\phi}$,
$(\partial_n\bm{\phi})\partial_n^2\bm{\phi}$, and
$\bm{\phi}\partial_n^3\bm{\phi}$ (which are
potentially dangerous for $\epsilon\geq 0$ according to power
counting) are redundant. A detailed RG analysis of the model with the
bulk and surface densities (\ref{eq:Lb}) and (\ref{eq:L1perp}) remains
to be done.

In the case of parallel surface orientation, it is possible to
investigate the ordinary transition without retaining the dependence
on $\lambda$ and $c$ \cite{DGR03}: In the limit $c\to c^*_{\rm
  ord}=\infty$ a Dirichlet boundary condition applies and the
dependence on $\lambda$ drops out (resides only in metric factors).
Hence one can set $\mathring{c}=\infty$ and $\mathring{\lambda}=0$,
choosing from the outset Dirichlet boundary conditions for the bare
theory. The critical exponent $\beta_1$ of the surface order parameter
$\bm{\phi}^{\mathfrak{B}}(\bm{r})=
\bm{\phi}(\bm{r},0)$ follows via the boundary operator
expansion
\begin{equation}
  \label{eq:boepar}
  \bm{\phi}(\bm{r},z)\mathop{\approx}_{z\to
    0}C(z)\,\partial_n\bm{\phi}\;, \quad  
C(z)\sim z^{\Delta[\partial_n\phi]-\Delta[\phi]}\;,
\end{equation}
giving $\beta_1^{\rm ord}/\nu_{L2}=\Delta[\partial_n\phi]$. Hence one
must study multi-point cumulants involving an arbitrary numbers of
fields $\bm{\phi}$ and boundary operators
$\partial_n\bm{\phi}$.  This strategy was followed in
Ref.~\cite{DGR03} and utilized to determine the critical index
$\beta_1^{\rm ord}$ to $O(\epsilon^2)$ for general $0\leq m\leq 6$.
The $\epsilon^2$ term involves a further single integral $j_1(m)$,
which again could be computed analytically for $m=0,2,6$, though only
numerically for other values. All other surface exponents of the
ordinary transition can be expressed in terms of a single one, e.g.\ 
$\beta_1^{\rm ord}$ and four independent bulk indices. The form
(\ref{eq:lamexp}) of the $\epsilon$ expansion, with $m$-independent
$O(\epsilon)$ terms, also applies to these surface exponents.
Furthermore, for $m\to 0$ their expansions to $O(\epsilon^2)$ turn
into the known ones \cite{Die86a,DD8081} of the standard semi-infinite
$\phi^4$ model. The ${d=3}$ estimates one obtains from these
$\epsilon$ expansions in the uniaxial scalar case ${m=n=1}$ (e.g.,
$\beta_1^{\rm ord}\simeq 0.68\ldots 0.7$) agree reasonably well with
recent Monte Carlo results for the ANNNI model \cite{Ple02}, which
gave $\beta_1^{\rm ord}=0.687(5)$.

The special transition is harder to analyze because the
$\lambda$-dependence must be retained, though $c$ can be set to its
fixed-point value $c^*_{\rm sp}=0$. A recent one-loop analysis
\cite{DR04} showed that $\beta_1^{\rm sp}$ agrees with the bulk
exponent $\beta_L$ to $O(\epsilon)$ and that the crossover exponent
$\Phi$ associated with $c$ becomes $m$-dependent already at
$O(\epsilon)$. According to recent Monte Carlo results
\cite{PH01,Ple02}, $\beta_1^{\rm sp}=0.23(1)$ and $\beta_L=0.238\pm
0.005$. Thus the difference $\beta_1-\beta_L$ seems to be small indeed.

Returning briefly to the case of perpendicular surface orientation,
let me conclude with a---hopefully educated---guess concerning the
ordinary transition. I expect that the asymptotic behavior at this
transition is described by a theory that obeys the boundary conditions
${\bm{\phi}^{\mathfrak B}=\partial_n\bm{\phi}=0}$. The
critical exponent $\beta_1$ in this case should follow from the
boundary operator expansion
\begin{equation}
  \label{eq:boeperp}
   \bm{\phi}(\bm{r},z)\mathop{\approx}_{z\to
    0}C_\perp(z)\,\partial_n^2\bm{\phi}\;, \quad  
C_\perp(z)\sim z^{\left(\Delta[\partial_n^2\phi]-
    \Delta[\phi]\right)/\theta}\;, 
\end{equation}
and be given by $\beta_1^{\rm ord}/\nu_{L2}=\Delta[\partial_n^2\phi]$.

\section*{Acknowledgments}
I thank Anja Gerwinski, Sergej Rutkevich, Mykola Shpot, and Royce K.P.
Zia with whom I had the pleasure to collaborate on different parts of
the work reported here. I also gratefully acknowledge the partial
support provided by the Deutsche Forschungsgemeinschaft (DFG) via the
grant Di-378/3.


\begin{thebibliography}{10}
\bibitem{aging}
Henkel has made analogous predictions for scaling functions one encounters in
  the study of relaxation phenomena. These predictions are at variance with the
  recent results to order $\epsilon^2$ of Ref.~\cite{CG02b} for the so-called
  model A.
\end{thebibliography}
\end{document}